\documentclass[aps,prl,twocolumn,groupedaddress]{revtex4-1}
\usepackage{amsmath}
\usepackage{amssymb}
\usepackage{graphicx}
\usepackage{subfigure}

\begin{document}


\title{Shear-wave-induced softening and simultaneous compaction in dense granular media through acoustic lubrication at flow heterogeneities}


\author{Charles K. C. Lieou}
\email[]{clieou@utk.edu}
\affiliation{Department of Nuclear Engineering, University of Tennessee, Knoxville, TN 37996}
\affiliation{Earth and Environmental Sciences, Los Alamos National Laboratory, Los Alamos, NM 87544, USA}
\author{Jerome Laurent}
\email[]{Present address: Univ\'ersit\'e Paris-Saclay, CEA, List, 91120 Palaiseau, France}
\affiliation{Institut Langevin, ESPCI Paris, Universit\'e PSL, CNRS, 75005 Paris}
\affiliation{LPMDI, Universit\'e Paris-Est Marne-la-Vall\'ee, 77454 Marne-la-Vall\'ee, France}
\author{Paul A. Johnson}
\affiliation{Earth and Environmental Sciences, Los Alamos National Laboratory, Los Alamos, NM 87544, USA}
\author{Xiaoping Jia}
\email[]{xiaoping.jia@espci.fr}
\affiliation{Institut Langevin, ESPCI Paris, Universit\'e PSL, CNRS, 75005 Paris}
\affiliation{Universit\'e Gustave Eiffel, 77454 Marne-la-Vall\'ee, France}


\date{\today}

\begin{abstract}
We report the simultaneous softening and compaction of a confined dense granular pack in acoustic resonance experiments. Elastic softening is manifested by a reduction of the shear-wave speed, as the wave amplitude increases beyond some threshold. No macroscopic rearrangement of grains or dilatancy is observed; instead, elastic softening is accompanied by a tiny amount of compaction on the scale of grain asperities. We explain these apparent contradictory observations using a theoretical model, based on shear transformation zones (STZs), of soft spots and slipping contacts. It predicts a linear shear stress-strain response with negligible macro-plastic deformation due to the small-amplitude acoustic oscillation. However, these waves reduce the interparticle friction and contact stiffness through the acoustic lubrication of grain contacts, resulting in an increase in the structural disorder or compactivity and softening of dynamic modulus. The compaction associated with this microscopic friction decrease is consistent with the prediction by an Ising-like correlation between STZs in the subyield regime.

\end{abstract}


\maketitle

Granular matter are ubiquitous in nature, and can demonstrate liquidlike or solidlike properties depending on the packing fraction and external load~\cite{jaeger_1996, durand_2000}. The viscoelastic and elastoplastic properties of dense granular media can be probed by bulk acoustic waves in a noninvasive fashion~\cite{liu_1992,liu_1993,jia_1999,makse_2004,jia_2004,hostler_2005,sen_2008,owens_2011,khidas_2012,knuth_2013}. Moreover, waves of sufficiently large amplitude are known to unjam the granular material via acoustic fluidization~\cite{melosh_1996,johnson_2005b} and reduce the aggregate elastic moduli, manifested by a decrease of sound speed through the granular pack~\cite{johnson_2005b,jia_2011}. 

Ref.~\cite{jia_2011} reported precise measurements of elastic softening and compaction induced by traveling compression waves and the subsequent healing~\cite{johnson_2005b,yoritomo_2020}, and proposed a microscopic model based on the Hertz-Mindlin theory of contacts for their observations within the mean-field framework. That study revealed that acoustic fluidization operates mainly through wave-induced shear contact stiffness weakening (via microslips), termed henceforth acoustic lubrication~\cite{brum_2018,leopoldes_2020}, rather than the acoustic pressure fluctuation (via contact opening) initially proposed~\cite{melosh_1996} and akin to the vibration fluidization~\cite{durand_2000}. Nevertheless, numerical simulations show that large-amplitude sound waves in a dense pack of elastic beads may change locally the contact network, on the mescoscopic scale, by a decrease of the coordination number~\cite{reichhardt_2015,lemrich_2017}, and create soft spots whose population increases with acoustic amplitude. 

In this work, we propose a model, based on shear transformation zones (STZs), to interpret qualitatively the observation of elastic softening and simultaneous compaction in confined granular glass bead packs, induced by \textit{shear waves} which cause much more pronounced irreversible sound-matter interactions than by compressional waves. This theoretical concept, originally developed for deformation in glassy, amorphous materials, provides new insights into the mesoscopic physics of nonaffine displacement and slippage of grains and the associated change in the force chain network at dynamic STZ heterogeneities~\cite{falk_1998,falk_2011,lemaitre_2002,lieou_2012,perchikov_2014,lieou_2015,lieou_2016,lieou_2017a,lieou_2017c}. 

Unlike the large-amplitude oscillatory shear measurements~\cite{perchikov_2014}, the STZ model gives rise to hysteresis curves whose encircled areas are narrow enough to resemble straight lines in the stress-strain phase space (i.e., almost linear stress-strain response), for shear waves of relatively small amplitude considered here. The elastic modulus can be inferred by the mean value over one oscillation cycle (see the Supplementary Material). Our experimental observation thus raises an important question: Is it possible to reconcile a linear STZ dynamics and a nonlinear acoustic response with a softening of shear wave velocity (dynamic modulus), as well as plastic deformation with a tiny amount of compaction on the scale of grain asperities? This situation is particularly relevant in dynamic earthquake triggering where the fault core may be impacted by various seismic waves ~\cite{johnson_2005b} from afar.

We will attribute these seemingly contradictory observations to an increase in disorder, manifested by an increase of an internal state variable or configurational temperature termed the \textit{compactivity}~\cite{mehta_1989} due to the decrease of interparticle friction and subsequent unlocking between neighboring grains through acoustic lubrication~\cite{jia_2011}. Such a mechanism may also allow the system to transition between metastable configurations, causing granular compaction~\cite{jaeger_1996,nowak_1998}.  

\begin{figure*}
\centering
\subfigure[~Experimental set-up]{\includegraphics[width=.35\textwidth]{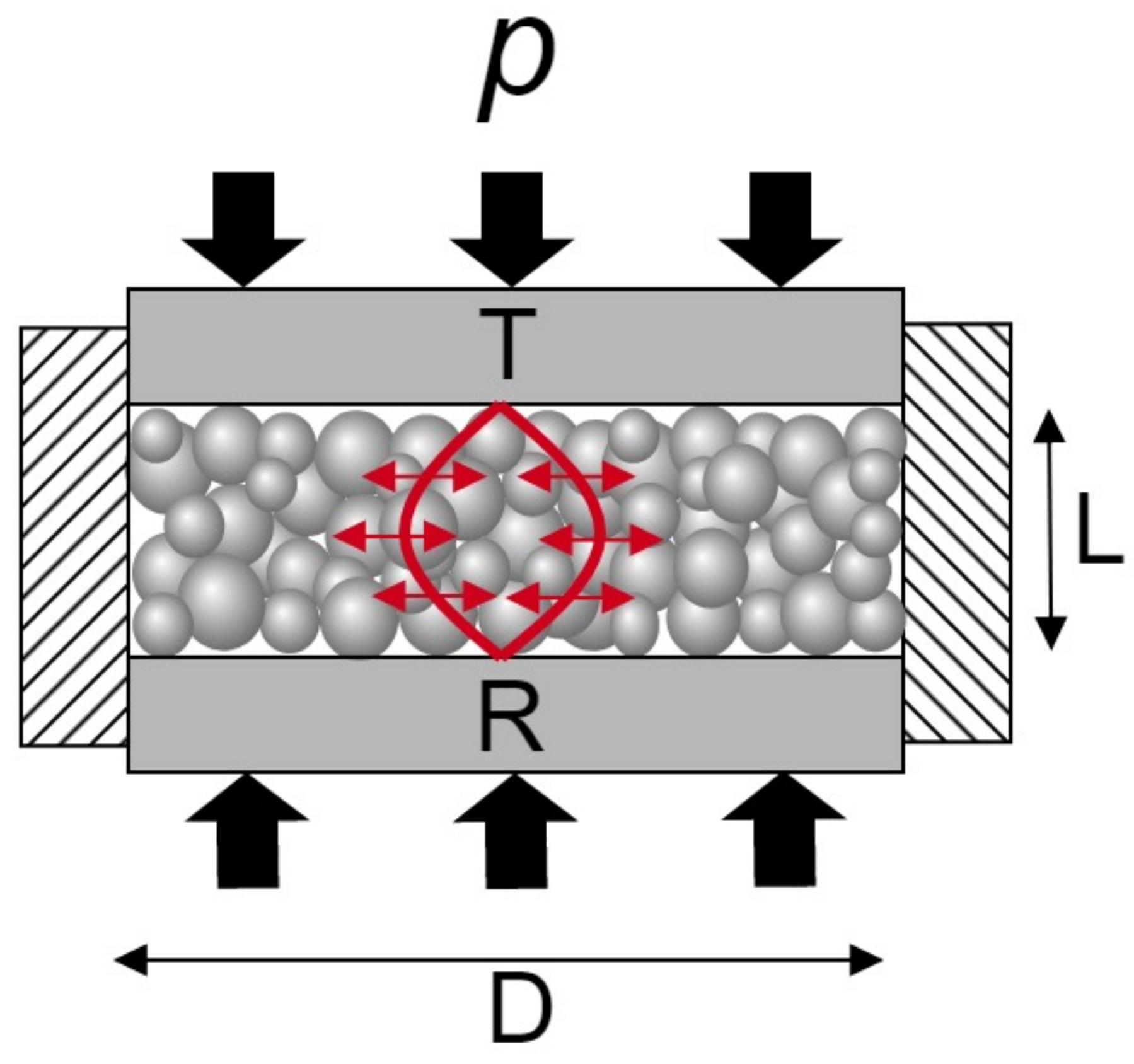}}
\subfigure[~Resonance curves for various amplitudes]{\includegraphics[width=.35\textwidth]{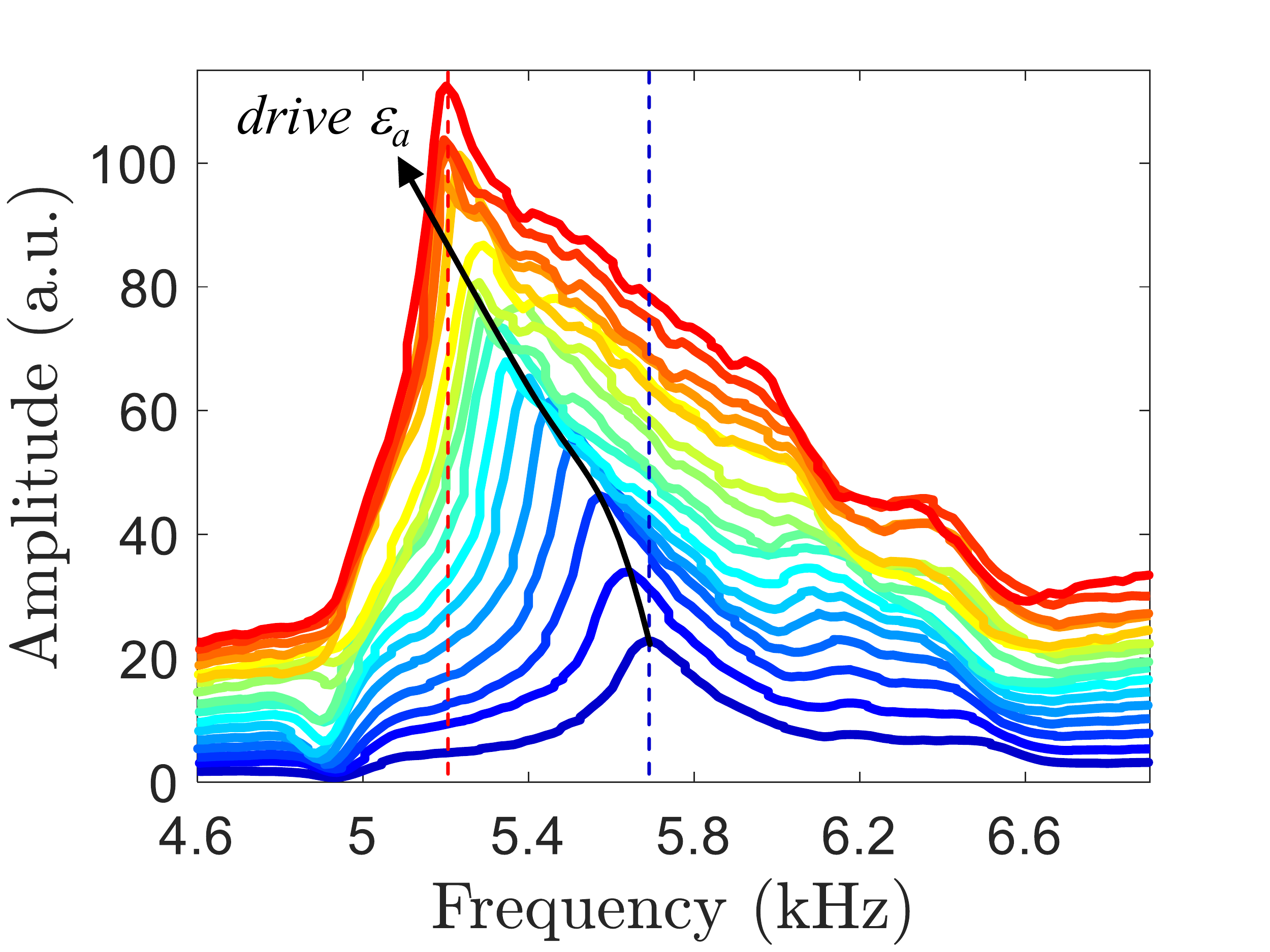}}
\subfigure[~Shear modulus weakening]{\includegraphics[width=.35\textwidth]{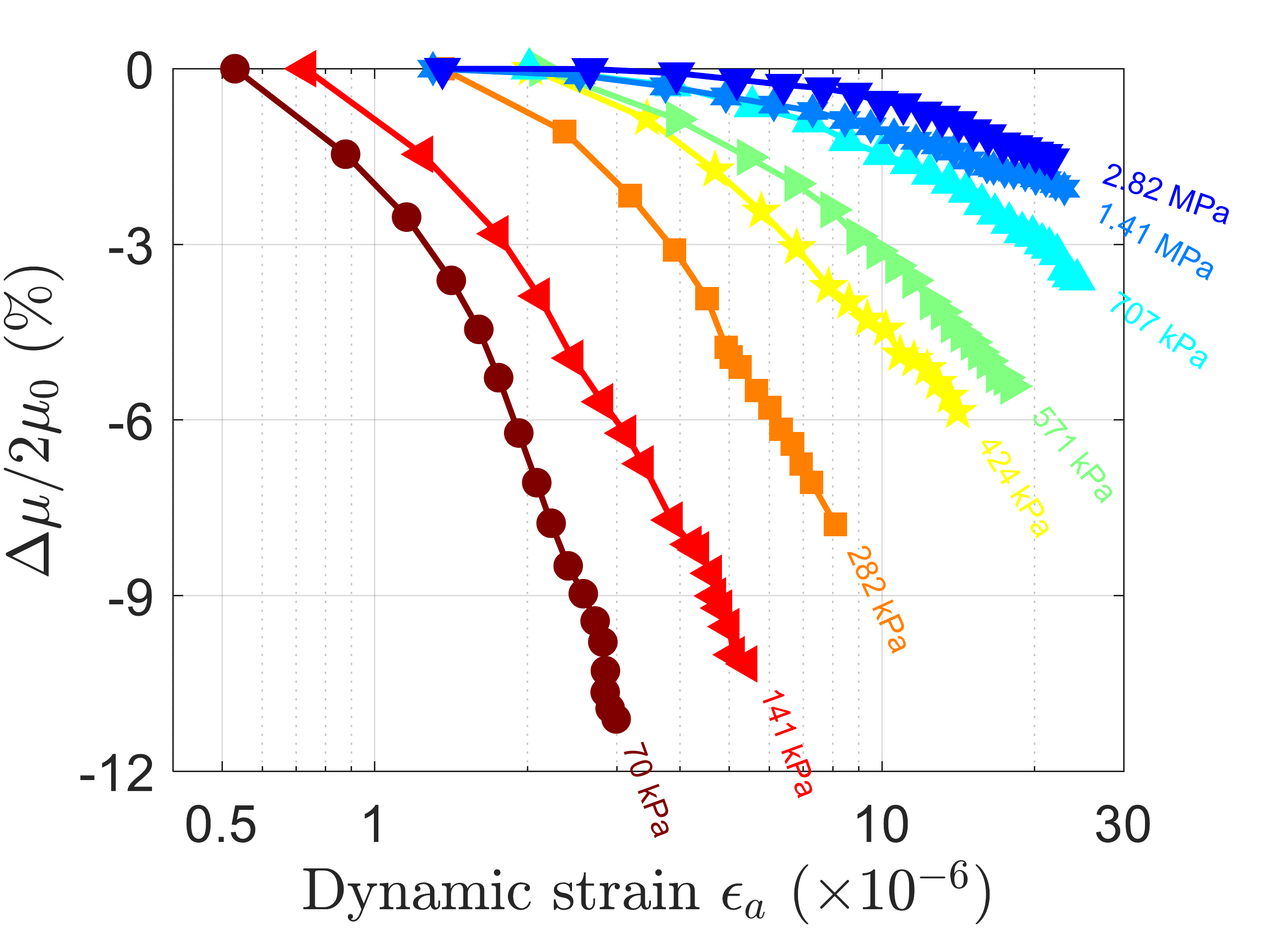}}
\subfigure[~Creep-like compaction]{\includegraphics[width=.35\textwidth]{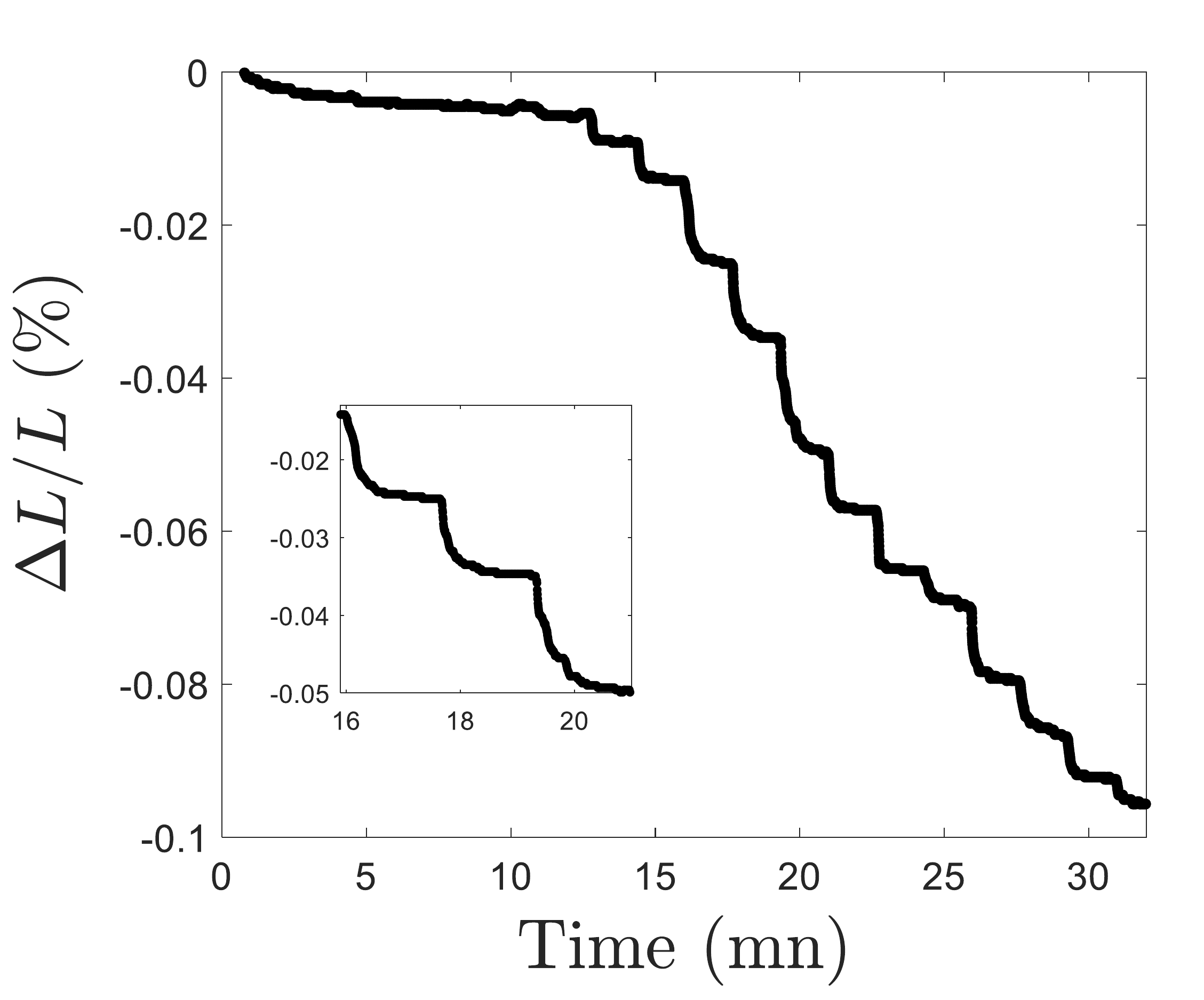}}
\caption{\label{fig:obs}Experimental observations. (a) Glass beads of average diameter $a = 0.6-0.8$ mm are confined in an oedometer cell at constant pressure $p$; a standing wave is established inside the cell with T the source transducer and R the receiver. (b) Resultant resonance curves of the fundamental shear mode for increasing drive amplitudes (indicated by the arrow) at $p = 140$ kPa. Downward shift of the resonance frequency indicates a reduction of the dynamic shear modulus. (c) Measured relative shear modulus softening compared to the unperturbed modulus $\mu_0$, for several values of the confining pressure $p$. (d) A step increase of the drive strain results in a small but detectable decrease of the packing thickness, compared to the shaking-induced granular compaction observed in ~\cite{nowak_1998}. A logarithmic-like relaxation following the abrupt microscopic compaction versus time is also visible (inset) under a given driving strain (here we do not attempt to quantify this dynamic behavior).}
\end{figure*}

\textit{Experiment.} Our granular materials consist of dry glass beads of diameter $a = 0.6 - 0.8$ mm and density $\rho_G = 2.4 \times 10^3$ kg m$^{-3}$, confined in a oedometer cell of diameter $D \approx 30$ mm and filled by rain deposition to a height $L \approx 18.5$ mm \cite{khidas_2012}. A normal load is applied on the bead pack across the top piston (piezoelectric transducer) using an electromechanical servo press, keeping the load constant within an error of $2\%$, corresponding to effective axial pressure $p$ ranging from 70 kPa to 2.8 MPa. Before acoustic measurements, one cycle of loading and unloading up to $p$ is performed on the granular packing in order to consolidate the sample and ensure repeatability of material preparation with a packing density $\phi = 0.63 \pm 0.01$. To optimize the propagation of coherent shear waves, a large shear transducer is used as a plane-wave source transmitting a continuous wave (Fig.~\ref{fig:obs}(a)). In general, the shear $v_s$ and compressional wave velocities $v_s$ and $v_p$ can be measured via traveling waves ~\cite{jia_1999} or the (fundamental) resonance frequency $f_r$ by $v_{p,s} = \lambda f_r$ with a wavelength $\lambda \approx 2 L$ for a resonator with rigid boundary conditions considered here~\cite{johnson_2005b}. To examine the nonlinear response, we vary the input voltage $V_{\text{input}}$ from 5 to 250 V, which correspond to a vibration displacement at the source transducer $u \approx$ 1 to 50 nm, calibrated by an optical interferometer. 

To construct a resonance curve, we sweep the frequency ranging from 1 to 10 kHz that contains the fundamental shear modes for roughly 120 s, and extract the time-averaged amplitude at each frequency interval. Figure \ref{fig:obs}(b) shows resonance curves in the glass bead pack under effective pressure $p = 140$ kPa. The graph shows a plot of detected amplitude versus sweeping frequency at progressively increasing drive amplitude. As the excitation amplitude is increased, the resonance frequency $f_r$ decreases which corresponds to a decrease in shear wave velocity $v_s$ and dynamic modulus $\mu$ (= $\rho_G v_s^2$). The broadening of resonance peak with increasing indicates a nonlinear frictional dissipation~\cite{brunet_2008}. Here, the induced modulus reduction in shear resonant modes is about $10\%$ (Fig. \ref{fig:obs}(c)), corresponding to a decrease in shear modulus of about $20\%$ over a drive strain range ($\epsilon_a$ = $\pi u/L$) of $10^{-6}$ to $10^{-5}$. This is about twice the P-wave induced modulus decrease over the same drive strain range~\cite{johnson_2005b}.

To investigate the irreversibility of the sound-matter interaction for increasing wave amplitude, we combine measurements of the resonance frequency downward shift and the packing density change (the sample height). Beyond a certain amplitude of driving depending on the confining pressure, the shear modulus weakening is accompanied by slight plastic deformation corresponding to a compaction of roughly 0.5 $\mu$m for one bead layer (Fig.~\ref{fig:obs}(d)). This characteristic scale suggests a micro-plastic behavior on the contact scale~\cite{jia_2011}, negligibly small compared to the macro-plastic rearrangement observed in shaking experiments on the grain scale~\cite{nowak_1998}. This finding highlights the sensitivity of the macroscopic elastic weakening on the tiny change of the contact network (via microslips) induced by sound waves~\cite{brum_2018} or weak vibration ~\cite{d'anna_2001} without visible macroscopic rearrangement of grains. Interestingly, such \textit{microscopic} compaction which undergoes a jump upon a step increase of the drive strain (with also a transient stress drop not shown) exhibits a logarithmic-like relaxation under external driving (inset of Fig.~\ref{fig:obs}(d)), reminiscent of those observed in the shaking-induced \textit{macroscopic} compaction ~\cite{nowak_1998}.

\textit{Theory: elastic softening through increase in compactivity}. Shaking ~\cite{durand_2000} and wave perturbation ~\cite{brum_2018} in weakly confined granular packs or simulations with frictionless particles \cite{reichhardt_2015} may give rise to soft spots, or STZ flow heterogeneities, where local slip and nonaffine displacement of the constituent particles -- discrete grains in the case of granular matter -- gives rise to nonzero strain. As a first approximation, we leave out an explicit description of the size of an STZ in its dynamics -- though a collectively slipping cluster may consist of dozens of particles \cite{abate_2007,daniels_2012} -- and assume a binary-cluster description based on whether a given soft spot is stable or unstable with respect to a deviatoric stress configuration with a given orientation. Denoting by $N$ the total number of grains, and $N_\pm$ the population of STZ heterogeneities in each of the two states, the STZ population size evolves according to the master equation
\begin{equation}\label{eq:master}
 \tau \dot{N}_{\pm} = {\cal R}(\pm \bar{s}) N_\mp - {\cal R}(\mp \bar{s}) N_\pm + \Gamma (N e^{-1 / \chi} - N_\pm ).
\end{equation}
The nonaffine part of the deviatoric strain rate due to STZ dynamics is
\begin{equation}\label{eq:v_pl}
 \dot{\gamma}^{\text{pl}}_{ij} = \dfrac{\epsilon_0}{\tau N} \dfrac{s_{ij}}{\bar{s}} ({\cal R}(\pm \bar{s}) N_\mp - {\cal R}(\mp \bar{s}) N_\pm) .
\end{equation}
Here, the time scale $\tau \propto \sqrt{\rho_G / p}$ is the characteristic time for a grain, driven by a pressure $p$ and otherwise not subjected to resistive forces, to move a distance equal to the typical grain size. $\bar{s} = \sqrt{(1/2) s_{ij} s_{ij}}$ is an invariant of the deviatoric stress tensor $s_{ij}$, which influences the rate ${\cal R}(\pm \bar{s}) / \tau$ at which forward and backward rearrangements occur; ${\cal R}(\pm \bar{s})$ is simply proportional to the probability per unit time of nonaffine grain displacements exceeding some threshold value \cite{falk_1998} of the order of a small fraction of the grain size. $\epsilon_0$ is an order-one parameter that measures the size of the STZ core relative to the typical grain size. $\Gamma = (\tau s_{ij} \dot{\gamma}_{ij}^{\text{pl}} ) / (\epsilon_0 s_0 \Lambda)$, where $\Lambda = (N_+ + N_-)/N$ is the STZ density, and $s_0$ is a stress scale, is proportional to the input work rate $s_{ij} \dot{\gamma}_{ij}^{\text{pl}}$; it is the rate at which the STZ density approaches the nonequilibrium steady-state value $\Lambda^{\text{ss}} = 2 e^{-1 / \chi}$ controlled by a structural temperature $\chi$ termed the compactivity (here dimensionless)~\cite{lieou_2012,lieou_2015} that measures the amount of configurational disorder in the granular pack. Since we focus on the nonequilibrium steady state defined by the standing wave excitation, $\Lambda = \Lambda^{\text{ss}}$ henceforth. Using $m \equiv (N_+ - N_-) / (N_+ + N_-)$, ${\cal C}(\bar{s}) \equiv ({\cal R}(\bar{s}) + {\cal R}(- \bar{s}) )/2$, and ${\cal T}(\bar{s}) \equiv ({\cal R}(\bar{s}) - {\cal R}(- \bar{s}) ) / ({\cal R}(\bar{s}) + {\cal R}(- \bar{s}) ) $, we simplify Eqs.~\eqref{eq:master} and~\eqref{eq:v_pl} to give 
\begin{eqnarray}
 \label{eq:m} \tau \dot{m} &=& 2 {\cal C}(\bar{s}) ({\cal T}(\bar{s}) - m) \left(1 - \dfrac{m \bar{s}}{s_0} \right) ; \\
 \label{eq:v_pl1} \tau \dot{\gamma}^{\text{pl}}_{ij} &=& 2 \dfrac{s_{ij}}{\bar{s}} \epsilon_0 e^{-1 / \chi} {\cal C}(\bar{s}) ({\cal T}(\bar{s}) - m ).
\end{eqnarray}

Wave velocity reduction of a traveling wave through granular matter, and the downward shift of a resonance peak in a confined granular medium, have been shown to be due to the same physical mechanism \cite{jia_2011}, exhibiting the same degree of softening for identical acoustic strain amplitude. However, wave speed reduction is directly tied to the reduction of effective elastic modulus or contact stiffness, and is therefore more fundamental. Thus, we analyze mathematically the reduction of wave speed and elastic modulus associated with traveling waves driven at a frequency equal to the measured resonance peaks. We also restrict ourselves to shear waves for the rest of this Letter; the result for compression waves is similar and will be reported in the Supplementary Material. Suppose that the traveling shear wave propagates in the $x$-direction and the displacement is in the $y$-direction. Then the shear strain is $\gamma = \partial u / \partial x = 2 \gamma_{xy}$, and the shear stress is $s = s_{xy} = s_{yx}$. Linearizing Eqs.~\eqref{eq:m} and \eqref{eq:v_pl1} around the small, oscillating quantities $s$ and $m$, and keeping in mind that ${\cal T}(s) = \tanh \left( \Omega s / \chi \right)$ with $\Omega = \epsilon_0 / (\epsilon_Z p)$ where $\epsilon_Z$ is the void volume of an STZ \cite{lieou_2012}, we find $\tau \dot{m} = 2 R_0 (\Omega s/\chi - m)$, and $\tau \dot{\gamma}^{\text{pl}} = 4 R_0 \epsilon_0 e^{-1 / \chi} ( \Omega s/\chi - m)$
with $R_0 \equiv {\cal C}(s = 0)$. Note that we are in the subyield regime  as $s \ll s_0$, unlike the post-yield regime as in, for example, Refs.~\cite{lieou_2012,lieou_2014a,lieou_2014b,lieou_2015,lieou_2016,lieou_2017a}, where we could use the approximation $\dot{m} = 0$; here $m$ is oscillating back and forth (see the Supplementary Material for a discussion of this point). Combine these with the statement of linear elasticity, $\dot{s} = \mu_0 (\dot{\gamma} - \dot{\gamma}^{\text{pl}})$, where $\mu_0$ is the unperturbed shear modulus, and the continuum version of Newton's second law, $\rho_G \ddot{u} = \partial s / \partial x + F$, where $F$ is the force field set up by the transducer. The steady-state oscillations suggest the ansatz $u = \hat{u} e^{i (k x - \omega t)}$, where $k = \omega / v$ is the wavenumber chosen by the resulting wave speed $v$, and similarly for $s$, $m$, and $F$. 

There is also an equation for describing the temporal evolution of the compactivity, of the form $\dot{\chi} \propto - s \dot{\gamma}^{\text{pl}} \chi  / \chi^{\text{ss}} + s\dot{\gamma}^{\text{pl}}$, that describes a competition between local optimization of grain packings (logarithmic relaxation; first term) and the creation of new, random loose packings by the mean flow in the post-yield regime (dilatancy; second term)~\cite{lemaitre_2002}. For small-amplitudes waves, $\chi$ is of order $\hat{u}^2$, so it suffices to use a fixed $\chi$ dependent on the amplitude; see Eq.~\eqref{eq:chi} below. Then we arrive at a direct relationship between the forcing amplitude $\hat{F}$ and the response amplitude $\hat{u}$: $\rho_G \omega^2 \hat{u} = A(v) \hat{F}$, where
\begin{equation}\label{eq:A}
 A(v) \equiv \dfrac{(\beta_s - i \omega) v^2}{(v_{s0}^2 \alpha - v^2 \beta_s) - i \omega (v_{s0}^2 - v^2) }
\end{equation}
is the response function, with $v_{s0} = \sqrt{\mu_0 / \rho_G}$ being the wave speed in the unperturbed limit, and $\alpha \equiv 2 R_0 / \tau$, $\beta_s \equiv \alpha (1 + 2 \mu_0 \epsilon_0 \Omega e^{-1 / \chi} / \chi)$.

\begin{figure}
\includegraphics[width=.4\textwidth]{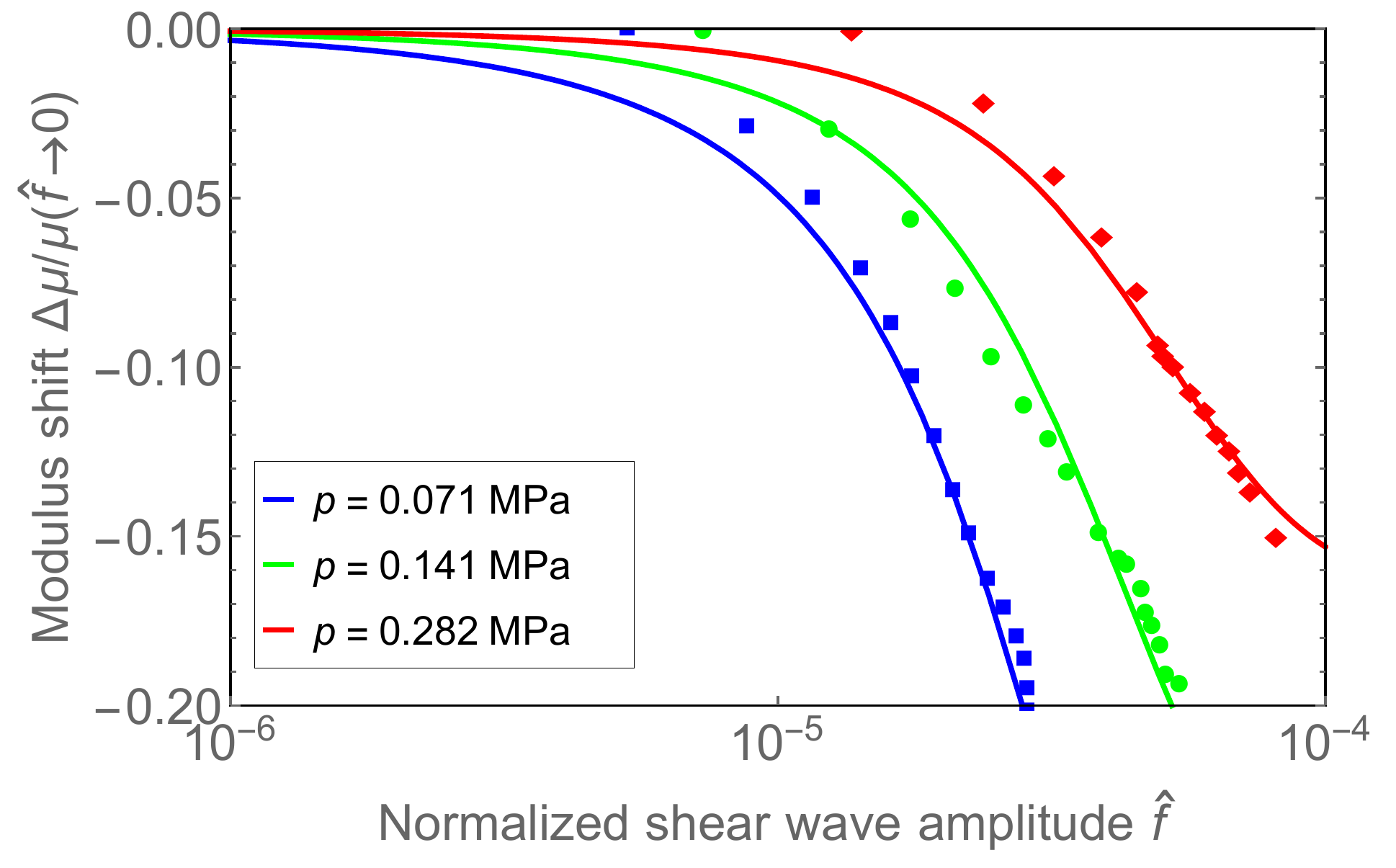}
\caption{\label{fig:softening}Measured relative shear modulus shift $\Delta \mu / \mu (\hat{f} \rightarrow 0)$, as a function of input amplitude, at several different pressures $p$. The data points are the experimental measurements, while the solids curves are fits obtained from Eqs.~\eqref{eq:A} and \eqref{eq:chi}.}
\end{figure}

In the steady state, the traveling wave at frequency $\omega$ propagates at a velocity that maximizes the amplitude $\hat{u}$ and hence $|A (v)|$. Thus
\begin{equation}
 \dfrac{v_s}{v_{s0}} = \sqrt{\dfrac{\alpha^2 + \omega^2}{\alpha \beta_s + \omega^2}},
\end{equation}
for the $S$-wave velocity $v_s$, corresponding to reduced shear modulus
 $\mu/\mu_0 = (v_s/v_{s0})^2$. The maximization principle for $|A (v)|$ arises from the excitation of eigenmodes; the response must be finite because of the STZ-induced damping. In the limit of low STZ density or small $\chi$, $\beta_{s} \rightarrow \alpha$, so that $v_s \rightarrow v_{s0}$. An increase of the amplitude of the wave perturbation generates more soft spots and is expected to increase $\chi$, thereby lowering the wave propagation speed $v_s$ and elastic moduli $\mu$ below the unperturbed value.

In the absence of other driving modes such as unidirectional rate-dependent shear, one expects that $\chi$ is related to the kinetic energy of the grains (i.e., mechanical noise)~\cite{lieou_2014b}. However, such a mechanism akin to the picture of effective temperature (vibrational) \cite{durand_2000,daniels_2012} would increase the volume of granular packing, which is contrary to the experimentally observed compaction.  

We follow here the scenario proposed in Ref.~\cite{jia_2011} in which the modulus softening is directly related to the shear acoustic lubrication of the contact area between grains, or microslipping at contacts that were originally stuck. This contact area is the state variable in the well-known rate-and-state friction law \cite{baumberger_2006}. Despite the different physical meaning of our state variable here, namely, the compactivity $\chi$, we posit in the steady state a dependence of $\chi$ on the acoustic amplitude thanks to a increase in \textit{fluidity} for the configurational change of packings by the acoustic lubrication on the microscopic scale at grain contacts, without rearrangement of grains or visible change of their relative topological positions. This is associated with an increase in the configurational disorder with amplitude. Without loss of generality, we adopt a functional form
\begin{equation}\label{eq:chi}
 \chi(\hat{f}) = \chi_0 + \chi_1 \tanh \left[ \left( \dfrac{p_0}{p} \right)^{1/6} \hat{f} \right],
\end{equation}
where $\hat{f}$ is a normalized ``deviatoric'' drive amplitude, and $\chi_0$ and $\chi_1$ control the permissible range of values of the compactivity, is found to provide a reasonably good fit to the experimental data. Note that if $\chi$ were independent of the drive amplitude, as in the small-amplitude treatment of oscillatory shear in \cite{perchikov_2014}, there can be no elastic softening at the amplitudes probed in the present experiments. Figure \ref{fig:softening} shows the measured modulus shift $\Delta \mu / \mu (\hat{f} \rightarrow 0)$, where $\Delta \mu = \mu - \mu (\hat{f} \rightarrow 0)$, at several different pressures $p$. The critical strain amplitude, at which the density of soft spots $\Lambda^{\text{ss}} = 2 e^{-1 / \chi}$ proliferates and causes appreciable softening, is found to be $\hat{f}_{\text{crit}} \sim 10^{-5}$. We used the parameters $\epsilon_0 = 1.5$, $\epsilon_Z = 0.5$ (as is typical in prior STZ descriptions of granular matter \cite{lieou_2012,lieou_2015,lieou_2016}), $\chi_0 = 0.065$, $\chi_1 = 0.02$, and $p_0 = 2.2 \times 10^{22}$ GPa.

\begin{figure}
\includegraphics[width=.4\textwidth]{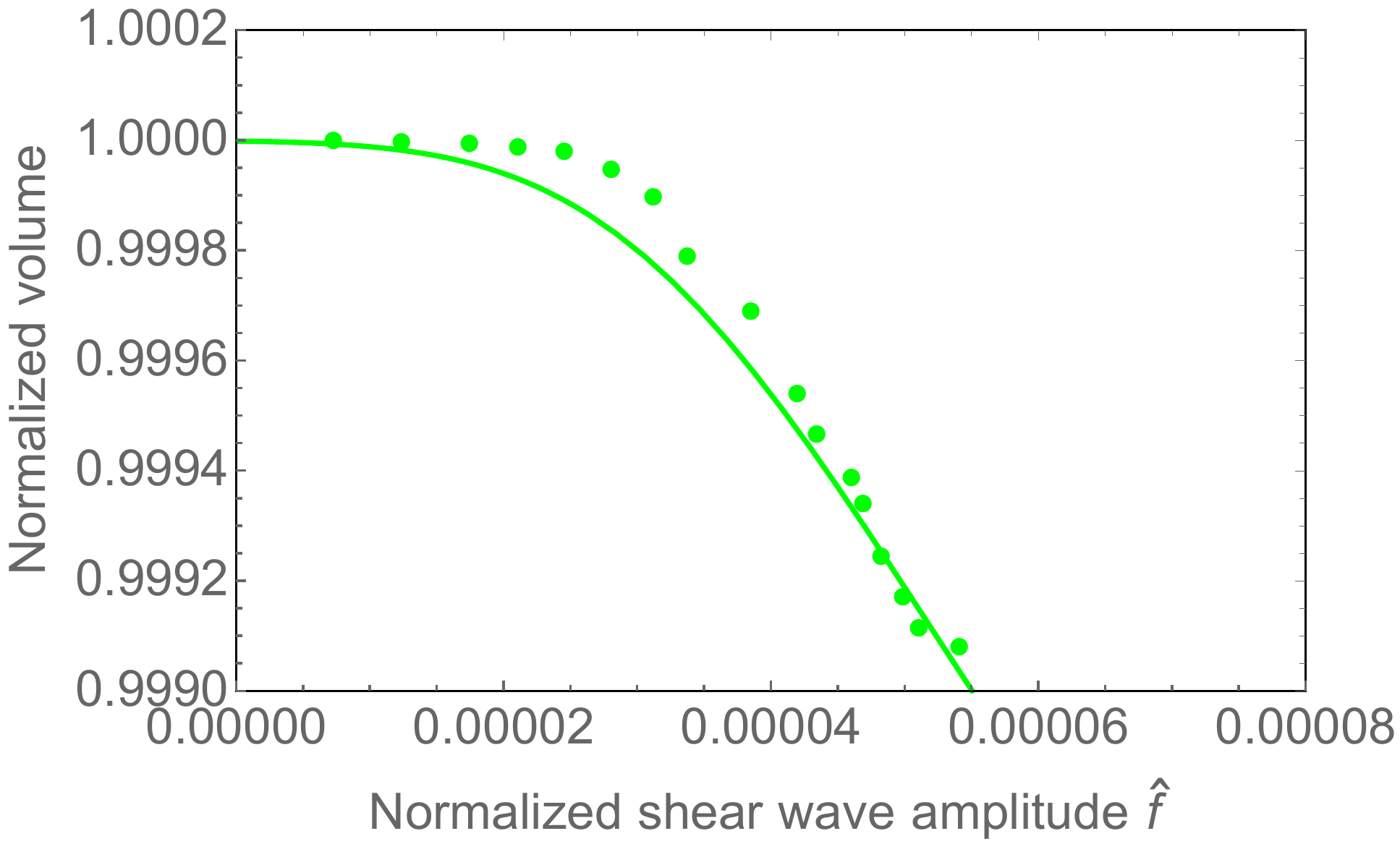}
\caption{\label{fig:compaction}Compaction from the initial volume, induced by shear waves, as a function of the amplitude $\hat{f}$, for $p = 0.141$ MPa.}
\end{figure}

\textit{Compaction in conjunction with softening.} The experimentally observed reduction of volume as a function of increasing acoustic strain amplitude, which causes progressive softening, is counterintuitive -- one might expect to observe dilatancy in conjunction with softening. This can be readily explained as follows. In addition to the bare volume $V_0$ of all grains and the average volume $v_Z$ of each STZ, there is an attractive interaction between the twofold-degenerate STZs which has been invoked to explain the Ising universality of glass-forming materials \cite{langer_2013}, for which granular matter is a prototype. STZs which are correlated with one another -- and are in the same state -- reduce the volume by some amount $v_M$. Mechanically, the correlation between STZ heterogeneities originates from the deformation at grain contacts. Thus the total volume $V$ occupied by the granular pack, divided by the bare volume $V_0$, is of the form
\begin{equation}
 \dfrac{V}{V_0} = 1 + \Lambda \epsilon_Z - \dfrac{\Lambda^2}{2} (1 + \hat{m}^2) \epsilon_M ,
\end{equation}
where $\epsilon_Z$ and $\epsilon_M$ are dimensionless quantities derived from $v_Z$ and $v_M$, and $\hat{m}$ is the amplitude of the oscillation of the orientational bias $m$, as in the above. The appearance of $\hat{m}^2$ follows from a straightforward treatment of the Ising symmetry. Figure \ref{fig:compaction} shows the normalized volume $V/V_0$ as a function of the drive amplitude $\hat{f}$, after the transient which we do not compute (i.e., the data points are taken near the end of each ``step'' in Fig.~1(d) where the volume ceases to decrease as a function of time for a given acoustic amplitude). In computing this curve we assumed that $\epsilon_Z = 0.5$, as above, and $\epsilon_M = 1.9 \times 10^7$, independent of pressure. It is seen immediately that above some critical strain amplitude -- same as that for the onset of modulus reduction -- the volume of the wave-perturbed granular packing decreases roughly linearly with increasing drive amplitude. For our choice of parameters, the degree of compaction remains small. Note that $\hat{m}$ itself is a function of the drive amplitude $\hat{f}$ and the parameters $\epsilon_Z$, $\epsilon_0$, $R_0$, $\chi_0$, $\chi_1$, and $p_0$; thus, with a single extra parameter $\epsilon_M$ that did not appear in the softening calculations above we are able to quantify compaction. Physically, the fact that compaction can occur in conjunction with softening can be attributed to the extra degrees of freedom at higher compactivity to explore possible configurations.


\begin{figure}
\includegraphics[width=.4\textwidth]{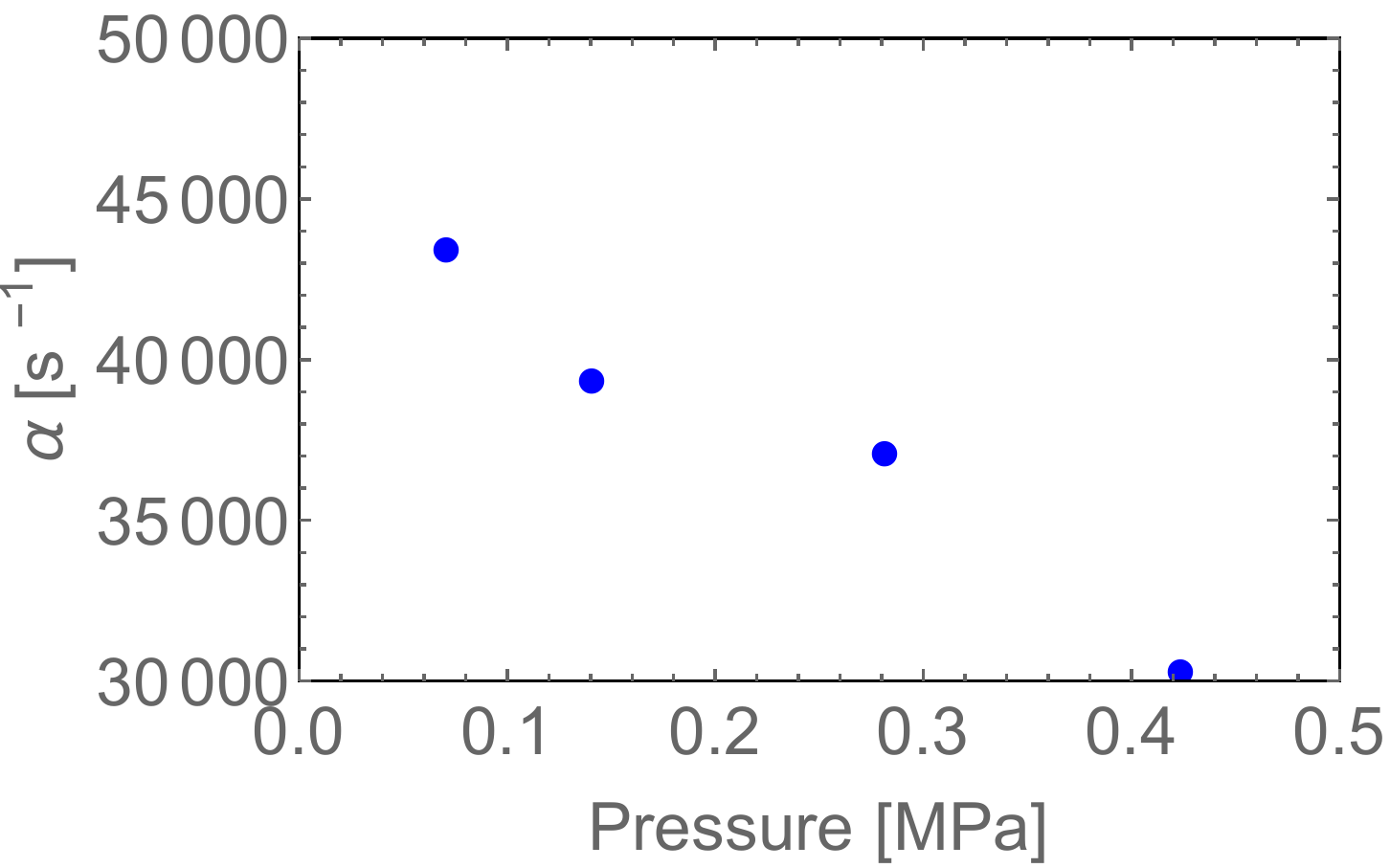}
\caption{\label{fig:alpha}Variation of STZ event rate $\alpha \equiv 2 R_0 / \tau$ (see text for the definition of $R_0$) with pressure. $\alpha$ decreases with increasing pressure $p$, suggesting slower slips, nonaffine displacement, and glassy dynamics, as one goes through the jamming transition.}
\end{figure}

\textit{Glassy dynamics.} Interestingly, to fit the experimental data we had to assume that the STZ event rate $R_0$ is a decreasing function of increasing pressure $p$; Fig.~\ref{fig:alpha} shows the calibrated values of the quantity $\alpha = 2 R_0 / \tau$ from the experiments. Because the asperity time scale $\tau$ scales as $p^{-1/2}$, the fact that $\alpha$ decreases with increasing pressure means that $R_0$ is also a decreasing function of pressure. If we assume that STZs describe the dynamical heterogeneities responsible for glassy phenomena in granular matter, as we did at the outset of this Letter, then the present result suggests that STZ activity appreciably slows down as both the pressure and packing fraction increase. This falls in line with Refs.~\cite{abate_2007,daniels_2012}, which observe in addition that slower flow heterogeneities increase in size, in effect moving cooperatively. If the latter also occurs in nonlinear resonance experiments in dense granular matter, this will point to tighter connections between softening and glassy dynamics than already established in this Letter. To this end, it will be important to find out, through future experiments and simulations that probe particle motion, whether cooperative rearrangements become more common as the pressure increases, through for example the $D_{\text{min}}^2$ criterion described in \cite{falk_1998}, or the four-point susceptibility in \cite{abate_2007}. Another avenue would be to measure whether and how the diffusion coefficient changes with pressure and drive amplitude; see for example \cite{langer_2012}.

In summary, we have attributed the softening and compaction in frictional granular matter driven by shear and compressional waves to microslips at the grain contact (microplasticity) and subsequent rearrangements of the contact network (logarithmic-like) without accompanying macroscopic grain motions (macroplasticity). The modeling based on the shear-transformation-zone framework allows us to uncover the relation between the state variable (i.e., the compactivity or configurational temperature) in the STZ model and the acoustic lubrication effect depicted by the contact model in ~\cite{jia_2011}. This lubrication-induced compaction on the microscopic scale is consistent with the prediction by an Ising-like correlation between STZs on the mesoscopic scale. We believe that this study will shed light on how the microscopic physics at grain contacts feed into glassy dynamics in the granular material, and help unravel the important connection between mesoscopic glassy dynamics such as activated-like hopping transition (logarithmic relaxation) \cite{d'anna_2001}, jamming and unjamming, and sound-matter interaction, which is also important for understanding natural hazards \cite{johnson_2005b,leopoldes_2020}.

\bibliography{prl_vf}

\end{document}



\title{Shear-wave-induced softening and simultaneous compaction in dense granular media through acoustic lubrication at flow heterogeneities -- Supplementary Material}


\author{Charles K. C. Lieou}
\email[]{clieou@utk.edu}
\affiliation{Department of Nuclear Engineering, University of Tennessee, Knoxville, TN 37996, USA}
\affiliation{Earth and Environmental Sciences, Los Alamos National Laboratory, Los Alamos, NM 87544, USA}
\author{Jerome Laurent}
\affiliation{Institut Langevin, ESPCI Paris, PSL, CNRS, 75005 Paris}
\affiliation{LPMDI, Universit\'e Paris-Est Marne-la-Vall\'ee, 77454 Marne-la-Vall\'ee, France}
\author{Paul A. Johnson}
\affiliation{Earth and Environmental Sciences, Los Alamos National Laboratory, Los Alamos, NM 87544, USA}
\author{Xiaoping Jia}
\email[]{xiaoping.jia@espci.fr}
\affiliation{Institut Langevin, ESPCI Paris, CNRS UMR 7587-1 rue Jussieu, 75005 Paris, France}
\affiliation{Universit\'e Paris-Est Marne-la-Vall\'ee, 77454 Marne-la-Vall\'ee, France}


\date{\today}

\begin{abstract}

\end{abstract}

\pacs{}

\maketitle


\section{Results for traveling pressure waves}

The results for the softening and compaction induced by traveling pressure waves are similar to those for shear waves described in the main manuscript, with some small differences. Equation (7) there remains intact for pressure waves, but with $\beta_s$ replaced by $\beta_p = \alpha (1 + \sqrt{3} M_0 \epsilon_0 \Omega e^{-1 / \chi} / \chi)$, where $M_0$ is the unperturbed P-wave modulus which correspondingly replaces $v_{s0}$ by $v_{p0}$. As such, the traveling P-wave propagates at a velocity
\begin{equation}
 \dfrac{v_p}{v_{p0}} = \sqrt{\dfrac{\alpha^2 + \omega^2}{\alpha \beta_p + \omega^2}} ,
\end{equation}
which corresponds to a reduced P-wave modulus
\begin{equation}
 \dfrac{M}{M_0} = \dfrac{\alpha^2 + \omega^2}{\alpha \beta_p + \omega^2}
\end{equation}
We use the functional form
\begin{equation}\label{eq:chi}
 \chi(\hat{f}) = \chi_0 + \chi_1 \tanh \left[ \left( \dfrac{p_0}{p} \right)^{1/6} \dfrac{\hat{f}}{\sqrt{3}} \right],
\end{equation}
to describe the variation of compactivity with strain amplitude under P-wave forcing; this differs from Eq.~(10) in the main manuscript by a factor of $1/\sqrt{3}$ in front of $\hat{f}$.

\begin{figure}
\includegraphics[width=.4\textwidth]{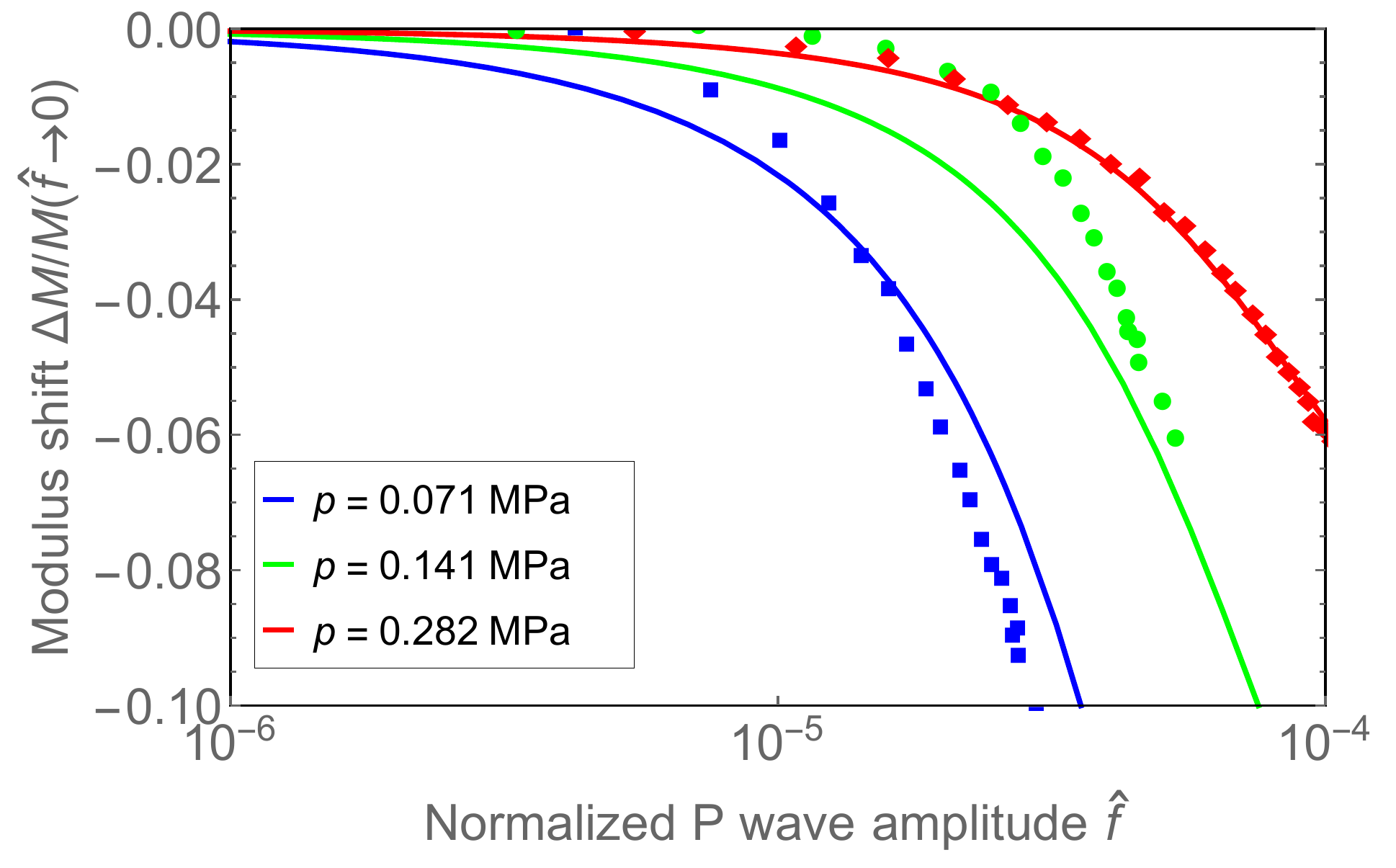}
\caption{\label{fig:softening}Measured relative modulus shift $\Delta M / M (\hat{f} \rightarrow 0)$, as a function of input amplitude, for the compressional mode, at several different pressures $p$. The data points are the experimental measurements, while the solids curves are fits obtained from Eq.~\eqref{eq:chi}. Note the difference in the vertical scales in comparison to Fig.~2 in the main manuscript, which shows that shear waves induce about twice as much modulus shift as pressure waves.}
\end{figure}

\begin{figure}
\includegraphics[width=.4\textwidth]{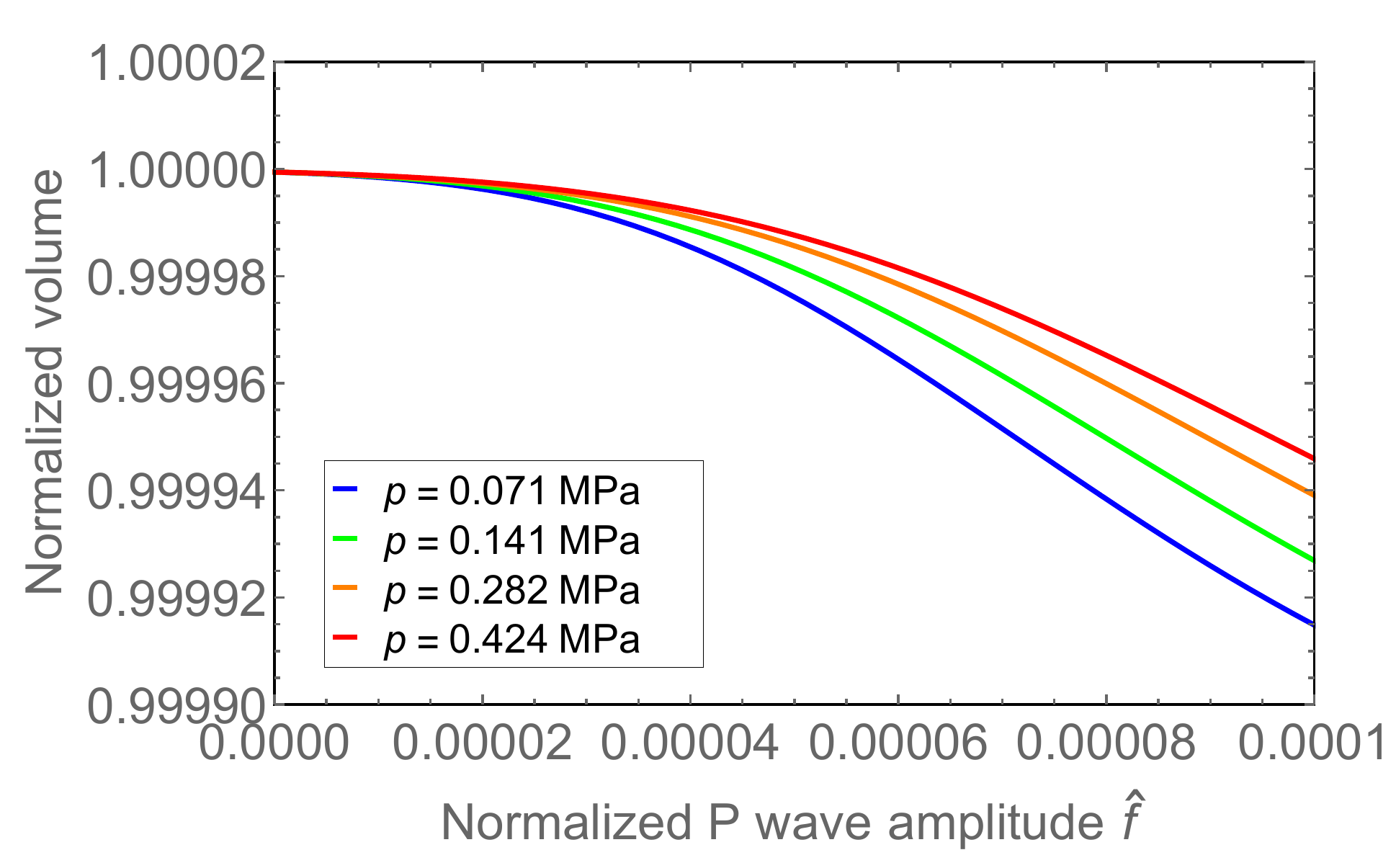}
\caption{\label{fig:compaction}Compaction from the initial volume, induced by pressure waves, as a function of the amplitude $\hat{f}$.}
\end{figure}

Figures \ref{fig:softening} and \ref{fig:compaction} show the results for the softening and compaction induced by P-waves, mirroring Figs.~2 and 3 in the main manuscript for shear waves. We used identical parameters as in the main manuscript, with one single exception: for P-waves we use $\chi_1 = 0.013$ instead of 0.02 for shear waves, to reflect the observation that the fraction of shear-induced resonance frequency shift is roughly twice that caused by standing pressure waves. This is reasonable because the resolved shear stress for the compressional mode is smaller than for the shear mode with the same stress magnitude, so that pressure waves causes fewer slips and produces fewer STZ soft spots than shear waves of equal amplitude. Interestingly, the compaction under P-wave perturbation is only about 1/10 that under S-wave perturbation. 

\section{Choice of parameters}

In computing the softening and compaction curves there are six adjustable parameters: $\epsilon_0$, $\epsilon_Z$, $\epsilon_M$, as well as the three parameters $\chi_0$, $\chi_1$, and $p_0$ that appear in the function $\chi (\hat{f})$. Of these, $\epsilon_M$ is the only one that appears in the compaction and not softening calculations. All others appears in both; notably, they appear in the quantity $\hat{m}$ which is computed by solving the linearized equations of motion.

The parameters $\epsilon_0 = 1.5$ and $\epsilon_Z = 0.5$ are regarded as fixed, based on prior STZ studies of granular matter (e.g., \cite{lieou_2012,lieou_2014a}). As usual, the values of other parameters are deduced systematically to fit the experimental data by slightly varying them one by one and observing the shift in the curves. Based on past experience with the STZ theory we start out by assuming that $\chi_0 \sim {\cal O} (0.1)$ and $\chi_1 \sim {\cal O} (0.01)$.

\section{Validity of linear approximation in the subyield regime}

Note that we have made the approximation $s \ll s_0$ in the evolution equation for the STZ orientational bias $m$ (Eq.~(5) in the main manuscript). This is not a valid approximation in flow problems (e.g., \cite{lieou_2012,lieou_2014a,lieou_2014b,lieou_2015,lieou_2016,lieou_2017a}), in which the granular material is unjammed and we are in the post-yield regime. However, insight gained from those problems do provide some insight about the present jammed, weakly-perturbed scenario. In flow problems, it is typically a good approximation to set $\dot{m} = 0$, so that 
\begin{equation}
 m = 
\begin{cases}
 {\cal T}(s), \quad \text{if $(s/s_0) {\cal T}(s) < 0$}; \\
 s_0 / s, \quad \text{if $(s/s_0) {\cal T}(s) \geq 0$}.
\end{cases}
\end{equation}
This shows that $s_0$ determines the yield stress of the unconsolidated granular pack. In the wave-perturbed granular pack, $m$ oscillates back and forth and $\dot{m} \ne 0$. But it is clear that we are in the subyield regime nevertheless, where the amplitude $\hat{m}$ of the $m$-oscillation increases with the amplitude $\hat{s}$ of the $s$-oscillation. The geometrical interaction between STZs, described by the term $\sim - \hat{m}^2$ in the expression for the volume (Eq.~(11)), gives rise to compaction with increasing strain amplitude. In contrast, in the post-yield regime in flow problems, the increase of the STZ density $\Lambda$ and the decrease of the orientational bias $m$ together give rise to dilatancy with increasing stress.

\section{Hysteresis in the linear approximation}

\begin{figure}
\includegraphics[width=.4\textwidth]{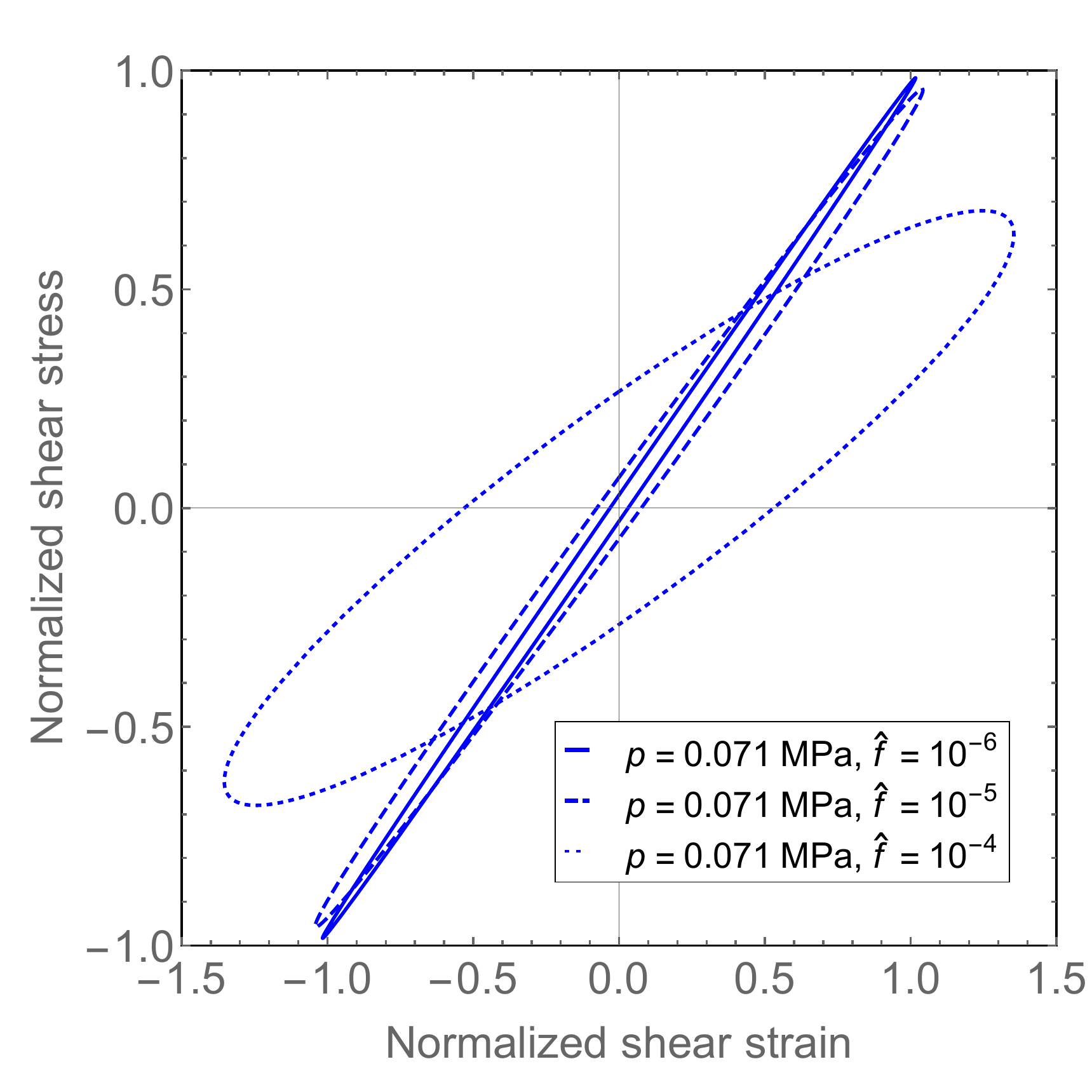}
\caption{\label{fig:s_hysteresis} Linear-approximation hysteresis curves of normalized shear stress $v_{s0} s(t) / (\mu_0 \omega \hat{u})$ versus normalized shear strain $v_{s0} \gamma(t) / (\omega \hat{u})$ for normalized drive amplitudes $\hat{f} = 10^{-6}$, $10^{-5}$, $10^{-4}$, at the pressure $p = 0.071$ MPa.}
\end{figure}

Figure \ref{fig:s_hysteresis} shows the post-transient shear stress $s(t)$ versus shear strain $\gamma(t)$ in the shear-wave experiment, both normalized with respect to their amplitudes, for several normalized drive amplitudes $\hat{f} = 10^{-6}$, $10^{-5}$, and $10^{-4}$, at the pressure $p = 0.071$ MPa. If the compactivity $\chi$ were independent of the driving acoustic amplitude ($\hat{f} = 0$), as in the small-amplitude treatment of oscillatory shear in \cite{perchikov_2014}, there can be no elastic softening at the amplitudes probed in the present experiments. Within the linearized approximation described in Eqs.~(5) and (6) in the main text, and the stress and displacement evolution equations discussed thereafter, the amplitudes $\hat{u}$, $\hat{\gamma}$, $\hat{s}$, and $\hat{F}$ are linearly related to one another, and the sinusoidal solution represents an ellipse in the $s$-$\gamma$ phase space. It is possible to reproduce hysteresis curves with sharp kinks such as seen in \cite{perchikov_2014}, and banana-shaped curves found in the literature (e.g., \cite{jia_2011}), once we move beyond the linear approximation. Still, our results still show salient features of hysteresis in dissipative materials. Namely, that the hystereis loop deviates further from a straight line through the origin as the drive amplitude increases, and that the slope of the semi-major axis of the hysteresis loop, which measures the shear modulus, decreases as one increases the amplitude. The result for the pressure-wave experiment is qualitatively similar.

\section{Softening in the zero-frequency and infinite-wavelength limits}

We showed in the main text that, with the ratio between the reduced and unperturbed shear moduli $\mu / \mu_0$ given by
\begin{equation}
 \dfrac{\mu}{\mu_0} = \dfrac{\alpha^2 + \omega^2}{\alpha \beta_s + \omega^2},
\end{equation}
where $\alpha \equiv 2 R_0 / \tau$, $\beta_s \equiv \alpha (1 + 2 \mu_0 \epsilon_0 \Omega e^{-1 / \chi} / \chi)$ and $\omega$ is the (angular) frequency of the shear wave, we concluded that the STZ event rate $R_0$ had to decrease with increasing pressure in order to arrive at a good fit with the pressure-dependent softening with increasing drive amplitude. A similar result holds for pressure waves). In the zero-frequency limit, or if $\alpha \gg \omega$, the softening behavior is independent of $R_0$. Said differently, we need a sufficiently high frequency $\omega \gtrsim R_0 / \tau$ for the pressure dependence to become apparent. Because $\tau = a \sqrt{\rho_G / p}$ is the inertial time scale and $\omega$ is related to the strain rate, one can think of the ``glassy slowing-down'' with increasing pressure as an inertial effect.

Perchikov and Bouchbinder \cite{perchikov_2014} studied the physically similar, yet somewhat different, limit of long wavelengths. Their paper was a proof-of-principles study of the response of generic amorphous materials to oscillatory shear, without taking into account pressure dependence. The wave speed for fixed frequency in the long-wavelength limit approaches infinity; it is therefore impossible to discuss the wave-induced softening measured by wave speed drop. Our present work differs in a significantly way; we combine the effects of waves and STZ heterogeneities and demonstrate the pressure-dependent softening and glassy behavior in granular matter in the high-frequency, short-wavelength regime.

\bibliography{prl_vf}